\documentclass[twocolumn,showpacs,prb]{revtex4}
\usepackage{graphicx}
\usepackage{dcolumn}
\usepackage{amsmath}
\usepackage{latexsym}
\begin{document}
\title{On the question of deconfinement in noncommutative Schwinger Model}
\author{Anirban Saha}
\altaffiliation{ani\_saha09@yahoo.co.in}
\affiliation{Department of Physics, Presidency College,\\86/1 College Street, Kolkata-700073, West Bengal, India.}
\author{Anisur Rahaman }
\altaffiliation{anisur.rahman@saha.ac.in}
\affiliation{Department of Physics, Durgapur Govt. College,\\
             Durgapur - 713214, Burdwan, West Bengal, India.}
\author{Pradip Mukherjee}
\altaffiliation{pradip@bose.res.in}
\altaffiliation{Also Visiting Associate, S. N. Bose National Centre for Basic Sciences, JD Block, Sector III, Salt Lake City, Calcutta -700 098, India and\\ IUCAA, Post Bag 4, Pune University Campus, Ganeshkhind, Pune 411 007,India}
\affiliation{Department of Physics, Presidency College,\\86/1 College Street, Kolkata-700073, West Bengal, India.}
\date{\today}
%\documentstyle[11pt]{article}
%\setlength{\topmargin}{0.2cm}
%\raggedbottom
%\abovedisplayskip=3mm
%\belowdisplayskip=3mm
%\abovedisplayshortskip=0mm
%\belowdisplayshortskip=2mm
%\setlength{\baselineskip}{16pt}
%\setlength{\evensidemargin}{0pt}
%\setlength{\oddsidemargin}{0in}
%\setlength{\parskip}{0.13cm}
%\setlength{\textwidth}{17truecm}
%\setlength{\textheight}{22cm}
%\baselineskip=20pt
%\renewcommand{\title}[1]{%
 %   \bigskip%
  %  \begin{center}%
   % \Large\bf #1%
    %\end{center}%
     % \vskip .2in}

%\renewcommand{\author}[1]{%
 %   {\begin{center}
  %  #1
  %  \end{center}}}
%\newcommand{\address}[1]{\vspace{-1.7em}\vspace{0pt}
 %   {\begin{center}
  %  \it #1
   % \end{center}}}
%\begin{document}
%\begin{titlepage}
%\title{On the question of deconfinement in noncommutative Schwinger Model}

%\author{Anirban Saha $\,^{\rm a,b}$,
%Anisur Rahaman $\,^{\rm c,d}$,
%Pradip Mukherjee\footnote{Also Visiting Associate, S. N. Bose National Centre 
%for Basic Sciences, JD Block, Sector III, Salt Lake City, Calcutta -700 098, India and\\ IUCAA, Post Bag 4, Pune University Campus, Ganeshkhind, Pune 411 007,India }
%$\,^{\rm a,e}$\ }
%\address{$^{\rm a}$Department of Physics, Presidency College\\
%86/1 College Street, Kolkata - 700 073, India}
%\address{$^{\rm c}$Department of Physics, Durgapur Govt. College,\\
%Durgapur - 713214, Burdwan, West Bengal, India}
%\address{$^{\rm b}$\tt ani\_saha09@yahoo.co.in}
%\address{$^{\rm d}$\tt anisur.rahman@saha.ac.in}
%\address{$^{\rm e}$\tt pradip@bose.res.in}

\begin{abstract}
\noindent
The $1+1$ dimensional bosonised Schwinger model with a generalized gauge invariant regularisation has been studied in a noncommutative scenario to investigate the fate of the transition from confinement to deconfinement observed in the commutative setting. We show that though the fuzziness of space time introduces new features in the confinement scenario,  it does not affect the deconfining limit. 
\end{abstract}

%\noindent {\bf PAC codes:}
%{\bf keywords}{Schwinger Model, Hamiltonian Analysis, Noncommutativity}

\maketitle

The $(1+1)$ dimensional Schwinger model has been extensively studied over the years \cite{old1, old2, old3, old4, old5, old6, old7, old8, old9, old10, old11, old12, old13, old14, old15, old16, old17, old18, ours} due to its significance in $(1+1)$ dimensional electrodynamics as well as due to its exactly integrable feature. 
One of the most interesting results following from the Schwinger model is the confining scenario of fermion \cite{old1, old3, old4}. Schwinger model with the usual gauge invariant regularization (GIR) leads to a massive photon where photon acquires mass via a kind of dynamical symmetry breaking and the fermion gets confined. The model is studied with different regularizations \cite{old10, old11, old14}. In fact ambiguity in the regularization which has played a crucial role in the confinement phenomena has been exploited several times \cite{old8, old9, old11, old12, harada} to explore various aspects of the model. The model is studied with a gauge non-invariant regularization too in \cite{old11, old14}. It was shown that fermions remained unconfined. However, photon acquired mass as usual. With a generalized gauge invariant regularization also the model is found to be consistent and physically sensible. This has a non-conventional limit where fermions enter into a deconfining phase \cite{old10}.
This transition from confinement to deconfinement is an extremely significant aspect and deserves to be studied in all its ramifications. 

In the recent times field theory formulated in noncommutative (NC) space time has gained considerable interest \cite{szabo} where the space time coordinates satisfy the algebra \cite{sny}
\begin{equation}
\left[x^{\mu}, x^{\nu}\right] = i \theta^{\mu \nu}
\label{ncgeometry}
\end{equation}
Investigation of the Schwinger model in NC space time has been initiated in \cite{ours} where we have studied the NC extension of the bosonised vector Schwinger model originating from the fermionic model under usual GIR. 
The introduction of noncommutativity gives rise to a novel 
background interaction which is linear in the NC parameter, signifying an interacting massive boson. Due to this background interaction the confinement phenomena can not be conclusively predicted. 
Naturally, study of the NC Schwinger model from different regularisation schemes is strongly suggested to elucidate the scenario of confinement. Specifically, it would be instructive to study the fate of fermion, i.e., whether it will appear as free or gets confined in a noncommutative space time configuration. This is all the more important because the fuzziness in the space time structure is considered to be effective in the Plank scale, where again the issues of confinement or deconfinement acquires cardinal significance.

 In view of the above perspective we propose to study the Schwinger model with a generalized GIR in a noncommutative setting. The commutative Schwinger model is described by the action 
\begin{eqnarray}
S_{F} = \int d^{2}x \left[ \bar\psi (i\partial\!\!\!\!/ -eA\!\!\!\!/)\psi
-{1\over 4}F^{\mu\nu}F_{\mu\nu}\right]
\label{faction}
\end{eqnarray}
where $\psi$ represents the fermionic field, $A_{\mu}$ stands for gauge field, $e$ has unit mass dimension and the Lorentz indices take the values $0, 1$.
Bosonisation of this model by integrating out the fermions requires some regularisation in order to remove the singularities in the fermionic determinant. Invoking a suitable generalized GIR \cite{old10} we obtain 
\begin{eqnarray}
S_{B}
% = \int d^{2}x {\mathcal{L}}
= \int d^{2}x \left[{1\over 2}\partial_\mu\phi\partial^\mu\phi
+\frac{e}{2}\epsilon_{\mu\nu}F^{\mu\nu}\phi
 + {\alpha\over 4}F_{\mu\nu}F^{\mu\nu} \right]
\label{baction}
\end{eqnarray}
The model, though looks simple, has a deeper meaning. Here $\alpha$ stands as a regularization ambiguity. In stead of mass-like regularization a kinetic energy-like regularization has been considered here. Unlike the former the kinetic energy-like regularization keeps the model gauge invariant which will be appropriate for the NC extension. We should mention here that $\mid \alpha \mid  \to \infty$ limit has a sharp contrast to $e = 0$ situation notwithstanding the fact that they look identical at first sight. A closer look reveals that $e = 0$ corresponds to a free electromagnetic theory. On the contrary $\mid \alpha \mid \to \infty$ limit is consistent with the occurrence of deconfinament as has been demonstrated from the calculation of quark--antiquark potential \cite{old10}. We will have occasions to further comment on this contrast arising out of noncommutativity later. For the time being we put $e = 1$ for simplicity and investigate the NC extension of (\ref{baction}) in this letter.

The NC field theories can be explored by several approaches which sometimes compliment each other \cite{pmas1, pmas2}. Thus one can think of the fields as operators carrying the realization of the basic algebra 
(\ref{ncgeometry}) or a conventional phase space may be used where the ordinary product is deformed. A particularly interesting scenario appears in case of the gauge theories where one can use the Seiberg--Witten (SW) type transformations \cite{SW, bichl, vic} to construct commutative equivalent models \cite{all1, all2, all3, all4} of the actual NC theories in a perturbative framework. In the present letter we adopt this approach.  
The noncommutative extension of the model (\ref{baction}) is then 
\begin{eqnarray}
\hat S = \int d^2x {\cal  L}_{NCBS} &=& \int d^2x \left[{1\over 2}\left(\hat{D}_\mu\star \hat\phi\right)
\star \left(\hat{D}^\mu\star\hat{\phi} \right) \right.  \nonumber\\
&&\left. +\frac{1}{2}\epsilon^{\mu\nu}\hat{\phi}\star\hat{F}_{\mu\nu}
 + {\alpha\over 4}\hat{F}_{\mu\nu} \star \hat{F}^{\mu\nu}\right]
\label{ncbaction}
\end{eqnarray}
with 
\begin{equation}
\hat D_{\mu}\star \hat \phi = \partial_{\mu} \hat \phi - i \left[\hat A_{\mu}, \phi\right]_{\star}
\label{covder}
\end{equation}
and $\star$ denotes that the ordinary multiplication is replaced by the star multiplication defined by
\begin{equation}
\hat \phi(x) \star \hat \psi(x) = \left(\hat \phi \star \hat \psi \right)(x) = e^{\frac{i}{2}
\theta^{\alpha\beta}\partial_{\alpha}\partial^{'}_{\beta}} 
  \hat \phi (x) \hat \psi(x^{'})\big{|}_{x^{'}=x.} 
\label{star}
\end{equation}
The action (\ref{ncbaction}) is invariant under the $\star$-gauge
transformation
\begin{equation}
\hat \delta_{\hat \lambda} \hat A_{\mu} = \hat D_{\mu} \star \hat \lambda, \qquad \hat \delta_{\hat \lambda} \hat \phi = -i \left[\hat \phi, \hat \lambda  \right]_{\star}
\label{starg}
\end{equation}
Note that in our model time space noncommutativity appears essentially. The issue of time space noncommutativity is a contentious one in the literature \cite{gomis, gaume, sei, bala, bala1}. Fortunately, in an order by order perturbative treatment consistent results have been obtained \cite{EW, EW1}. In an interesting work \cite{Daiy} the origin of the time space noncommutativity has been traced from a theory with only spatial noncommutativity to first order in the NC parameter, thereby ensuring regular behaviour of the derived theory. We have treated the $(1+1)$ dimensional Schwinger model in \cite{ours} which has also exhibited consistent results having the correct commutative limit. In the present work we are again confined to first order calculation. Thus the issue of time space noncommutativity is expected not to create any trouble.
To the lowest order in $\theta$ the explicit forms of the SW maps are known as \cite{SW,bichl, vic}
\begin{eqnarray}
\hat \phi &=& \phi - \theta^{mj}A_{m}\partial_{j}\phi \nonumber\\
\hat A_{i} &=& A_{i} - \frac{1}{2}\theta^{mj}A_{m}
\left(\partial_{j}A_{i} + F_{ji}\right)
\label{1stordmp}
\end{eqnarray}
Using these expressions and the star product (\ref{star}) to order $\theta$ in (\ref{ncbaction}) we get
\begin{eqnarray}
\hat S &\stackrel{\rm{SW \; map}}{=}& \int d^{2}x \left[\left\{ 1 + \frac{1}{2} {\rm{Tr}}\left(F \theta\right)\right\}{\mathcal{L}}_{c} - \left(F \theta\right)_{\mu}{}^{\beta}\partial_{\beta}\phi \partial^{\mu}\phi \right.\nonumber\\
&& \left. \qquad \quad - \frac{1}{2} \epsilon^{\mu \nu}\left(F \theta F\right)_{\mu \nu} \phi  - \frac{\alpha}{2}\left(F \theta F\right)^{\mu \nu}F_{\mu \nu}\right]\nonumber\\
\label{1storderac}
\end{eqnarray}
where ${\mathcal {L}}_{c}$ stands for the commutative Lagrangean (with $e = 1$)
\begin{eqnarray}
{\mathcal {L}}_{c} =  \frac{1}{2}\partial_{\mu}\phi \partial^{\mu}\phi + \frac{1}{2}\epsilon_{\mu\nu}F^{\mu\nu}\phi + {\alpha\over 4}F_{\mu\nu}F^{\mu\nu} 
\label{cL}
\end{eqnarray}

 We will now proceed with the Hamiltonian analysis of the model. The canonical momenta corresponding to the fields $\phi$ and $A_{\mu}$ following from the standard definitions are 
\begin{eqnarray}
\pi_{\phi} &=& \dot{\phi} + \theta F_{01}\dot{\phi}\nonumber\\
\pi^{0} &=& 0 \nonumber\\
\pi^{1} &=& - \alpha F_{01} + \phi + \frac{\theta}{2}\left({\dot{\phi}}^{2} - 
\phi^{\prime}{}^{2} - 3\alpha F_{01}^{2} \right)
\label{momenta}
\end{eqnarray}
The Hamiltonian density corresponding to the action (\ref{ncbaction}) is 
\begin{eqnarray}
%\int d^{2}x 
{\cal H}_{CEV} & = &
%\int dx 
\left[{\cal H}_{CS} - \frac{\theta}{2\alpha}
\left\{\pi^{1}\left(\phi^{\prime}{}^{2} - \pi_{\phi}{}^{2}\right) + \phi \left(\pi_{\phi}{}^{2} - \phi^{\prime}{}^{2}\right)\right\}\right.\nonumber\\
&&\left. 
+ \frac{\theta}{2\alpha^{2}}\left\{ \phi^{3} - \left(\pi^{1}\right)^{3} + \frac{3}{2}\phi \pi^{1}\left(\pi^{1} - \phi\right)\right\}\right]
\label{ncH}
\end{eqnarray}
where ${\cal H}_{CS}$ is given by 
\begin{equation}
{\cal  H}_{CS}= \frac{1}{2} \left[\pi_{\phi}{}^{2} - \frac{1}{\alpha}(\pi^{1}){}^{2} + \phi^{\prime}{}^{2} - \frac{1}{\alpha} \phi^{2} \right]
+ \pi^{1} A_{0}^{\prime} + \frac{1}{\alpha}\pi^{1}\phi
\label{cH}
\end{equation}
$\pi_{0} = 0$ in (\ref{momenta}) is a primary constraint, conserving it in time we get a secondary constraint $\pi^{1} = 0 $ and these two forms a first class set. We therefore choose the following two gauge fixing conditions 
\begin{eqnarray}
A_{0} & = & 0\nonumber\\
A_{1} & = & 0
\label{gfc}
\end{eqnarray}
in order to remove the gauge redundancy. The Hamiltonian in this reduced phase space is obtained by setting $ \pi_{0} = \pi^{1} = 0, \  A_{0} = A_{1} = 0 $.
\begin{eqnarray}
%\int dx 
{\cal H}_{RCEV} & = &
%\int dx 
\left[ \frac{1}{2} \left(\pi_\phi^2  + \phi^{\prime}{}^{2}-  \frac{1}{\alpha}  \phi^{2}\right) \right. \nonumber\\
&&  \left.- \frac{\theta}{2\alpha}\left\{\phi\pi_{\phi}^{2} - \phi \phi^{\prime2} \right\} + \frac{\theta}{2\alpha^{2}}\phi^{3}\right]
\label{ncHr}
\end{eqnarray}
In this situation the Dirac brackets \cite{dir}, and not the Poission brackets represent the simplectic structure. A straightforward calculation using the definition of the Dirac brackets shows that the brackets remain canonical. The Hamiltonian (\ref{ncHr}), along with these canonical brackets leads to the following first order differential equations
\begin{equation}
\dot\phi=\left(1 - \frac{\theta}{\alpha} \phi\right) \pi_{\phi} 
\label{EQ1}
\end{equation}
\begin{equation}
\dot\pi_{\phi} = \phi^{\prime \prime} + \frac{1}{\alpha}\phi + \frac{\theta}{2 \alpha} \left(\pi_{\phi}^{2} + \phi^{\prime}{}^{2} + 2 \phi \phi^{\prime \prime}\right) - \frac{3 \theta}{2 \alpha^{2}}\phi^{2}
\label{EQ2}
\end{equation}
The above two first order equations can be combine to the following second order equation 
\begin{equation}
\left( \Box - \frac{1}{\alpha} \right)\phi =  -\frac{\theta }{2\alpha} \left(\dot{\phi}^{2} - \phi^{\prime}{}^{2} + \frac{5}{\alpha} \phi^{2} \right) 
\label{box}
\end{equation}
The above equation, devoid as it is of the redundant degrees of freedom, contains the physical contents of the model (\ref{ncbaction}) to leading order in the NC parameter $\theta$. For a general $\alpha$ it exhibits a massive photon in interaction with a background which naturally agrees with our earlier finding \cite{ours, error}. However, the crux of the situation lies in the nonconventional limit $\mid\alpha\mid \to \infty$ which in one stroke eliminate both the background interaction as well as the mass term. We are then left with a massless boson which reveals a deconfined fermion.
It is indeed gratifying to observe that though the introduction of noncommutativity gives rise to a background interaction, in the large $\alpha$ limit this interaction is removed and the deconfining phase is retrieved.
Also note that this $\mid\alpha\mid \to \infty$ limit is very different from $e \to 0$ limit, a fact mentioned earlier in connection with the action (\ref{baction}). 

 We have discussed the bosonised Schwinger model obtained under a generalized gauge invariant regularisation (GIR) in a noncommutative (NC) setting in order to investigate the fate of the confinement scenario in the NC context. Using the standard $\star$-product formalism and invoking the appropriate Seiberg--Witten (SW) transformations the original NC model has been mapped to an equivalent commutative model. Performing a Hamiltonian analysis the basic field equations were derived in the reduced space. These field equations were combined to lead to a second order equation which contains the physical informations of the theory. The results of our calculations contain the regularisation parameter $\alpha$. For a general value of this parameter we obtained a massive boson in interaction with a background of NC origin, a result in agreement with \cite{ours}. Remarkably, in the large $\alpha$ limit this background interaction is eliminated along with the mass term which revealed the presence of a deconfined fermion.
 So $\mid\alpha\mid \to \infty $ is a limit which corresponds to the deconfining phase of the fermion. This unconventional limit was earlier studied in the commutative case \cite{old10} which revealed the same deconfinement scenario. Our results thus prove the existence of this deconfining phase in the NC perspective. This is a welcome result considering that the fuzziness of space time associated with the NC algebra is relevant near the Plank scale \cite{bert, rb, ani} where again the issue of confinement or deconfinement assumes special significance. 
%It would be sufficient to mention that the limit we consider will certainly be interesting and exciting in connection with the formation of QGP since QGP phase is considered as spontanious and natural near Plank scale after barryogenesis. Indeed, it is fair to say that the scope of probing the Plank scale is very limited at present both from theoretical and experimental framework.

%%%%%%%%%%%%%%%%%%%%%%%%%%%%%%%%%%
\section* {Acknowledgment}
%%%%%%%%%%%%%%%%%%%%%%%%%%%%%%%%%%
{AS wants to thank the Council of Scientific and Industrial Research (CSIR), Govt. of India, for financial support.}


\begin{thebibliography}{99}
\bibitem{old1} J.~Schwinger, Phys. Rev., {\bf 128} (1962) 2425.
\bibitem{old2} W. Thirring, Ann. Phys. {\bf 3} (1958) 91.
\bibitem{old3} J.~H.~Lowenstein, J.~A.~Swieca, Ann. Phys. {\bf 68} (1971) 172.
\bibitem{old4}S.~Coleman, R.~Jackiw, J.~A.~Swieca, Ann. Phys. (NY) {\bf 13} (1975) 267.
\bibitem{old5}S.~Coleman, Ann. Phys. (NY) {\bf 101} (1976) 239.
\bibitem{old6}H.~J.~Rothe, K.~D.~Rothe, J.~A.~Swieca, Phys.Rev {\bf D 19} {1979} 3020.
\bibitem{old7}K.~D.~Rothe, B.~Schroer, Nucl.Phys.{\bf B 185} (1981) 429.
\bibitem{old8}  R.~Jackiw and R. Rajaraman, Phys Rev.Lett. {\bf 54} (1985) 1219.
\bibitem{old9}R.~Banerjee, Phys.Rev.Lett.{\bf 56}(1986) 1889.
\bibitem{old10} G.~Bhattacharya,, A. Ghosh, P. Mitra, Phys. Rev. {\bf D50} 
(1994)4138.
\bibitem{old11} P. Mitra and A. Rahaman, Ann. Phys. (N.Y.) {\bf 249} (1996) 34
\bibitem{old12} A.Rahaman and P. Mitra, Mod. Phys. Lett. {\bf A11} (1996) 
2153.
\bibitem{old13} J.~Kijowski, G.~Rudolph, M.~Rudolph, Phys.Lett. {\bf B 419} (1998) 285,  [hep-th/9710003].
\bibitem{old14} A. Rahaman Int. Jour. Mod. Phys. {\bf A 19} (2004) 3013.
\bibitem{old15}S.~D\"{u}rr, C.~Hoelbling,  [hep-lat/0408039].
\bibitem{old16}H.~M.~Sadjadi, M.~Alimohammadi, Gen.Rel.Grav. 37 (2005) 1809, [gr-qc/0506030].
\bibitem{old17}R.~Shankar, G.~Murthy, [cond-mat/0508242].
\bibitem{old18}S.~Azakov, [hep-th/0511116].
\bibitem{ours}A.~Saha, A.~Rahaman, P.~Mukherjee; Phys. Lett. {\bf B 638} 292, 2006, [hep-th/0603050].
\bibitem{harada}K.~Harada, Phys. Rev. Lett. 64, 139 (1990), 
\bibitem{szabo} See R.~J.~Szabo, Phys. Rep. {\bf 378} (2003) 207 and the references therein.
\bibitem {sny}Heisenberg first suggested this idea which was later developed by Snyder; H.~S.~Snyder, Phys. Rev. {\bf 71} (1947) 38; {\it{ibid}} 72 (1947) 874.
\bibitem{pmas1}P.~Mukherjee, A.~Saha, Mod.Phys.Lett.{\bf{A21}} (2006) 821, [hep-th/0409248], 
\bibitem{pmas2}P.~Mukherjee, A.~Saha, [hep-th/0605123],
To appear in Mod. Phys. Lett. {\bf A}.
\bibitem{SW} N.~Seiberg and E.~Witten, JHEP {\bf 09} (1999) 032.
\bibitem{bichl} A.~A.~Bichl, J.~M.~Grimstrup, L.~Popp, M.~Schweda, R.~Wulkenhaar, [hep-th/0102103]. 
\bibitem{vic} V.~O.~Rivelles, Phys.Lett.{\bf B558} (2003) 191, [hep-th/0212262]
\bibitem{all1}O.~F.~Dayi, Phys.Lett. {\bf B 560} (2003) 239, [hep-th/0302074].
\bibitem{all2} S.~Ghosh, Nucl.Phys. {\bf B 670} (2003) 359, [hep-th/0306045].
\bibitem{all3}B.~Chakraborty, S.~Gangopadhyay, A.~Saha, Phys. Rev. {\bf D 70} (2004) 107707, [hep-th/0312292].
\bibitem{all4}S.~Ghosh, Phys.Rev.{\bf D70} (2004) 085007, [hep-th/0402029].
\bibitem{gomis} J.~Gomis, T.~Mehen, Nucl.Phys. {\bf B 591} (2000) 265, [hep-th/0005129].
\bibitem{gaume}L.~Alvarez~Gaume, J.~L.~F.~Barbon, R.~Zwicky, JHEP {\bf 05} (2001) 057, [hep-th/0103069].
\bibitem{sei}N.~Seiberg, L.~Susskind, N.~Toumbas, JHEP {\bf 06} (2000) 044, [hep-th/0005015].
\bibitem{bala}S.~Doplicher, K.~Fredenhagen, J.~Roberts, Phys. Lett. {\bf B 331} (1994) 39.
\bibitem{bala1}S.~Doplicher, K.~Fredenhagen, J.~Roberts, Comm. Math. Phys. {\bf 172} (1995) 187, [hep-th/0303037].
\bibitem{EW} D.~A.~Eliezer, R.~P.~Woodard, Nucl.Phys. {\bf{B 325}} (1989) 389.
\bibitem{EW1}T.~C.~Cheng, P.~M.~Ho, M.~C.~Yeh, {\it{ibid}} {\bf{B 625}} (2002) 151, [hep-th/0111160].
\bibitem{Daiy}O.~F.~Dayi, B.~Yapiskann, JHEP {\bf{10}} (2002) 022, [hep-th/0208043].
%\bibitem{Git}D.M. Gitman and I.V. Tyutin, Quantization of Fields with Constraints (Springer-Verlag, Berlin 1990)
\bibitem{dir} P.~A.~M.~Dirac, {\it Lectures on Quantum Mechanics}, (Yeshiva University Press, New York, 1964).
\bibitem{error}Note that there is a calculational error in the expression for the Hamiltonian in \cite{ours} which percolates to the subsequent equations derived from the Hamiltonian. For the correct expressions see erratum appended to [hep-th/0603050] (to appear in Phys. Letts. {\bf B}).
%\bibitem{ref1} O.~F.~Dayi, A.~Jellal, J.Math.Phys. {\bf{43}} (2002) 4592, [hep-th/0111267].
%\bibitem{ref2}F.~G.~Scholtz, B.~Chakraborty, S.~Gangopadhyay, A.~Ghosh~Hazra, Phys. Rev. {\bf{D 71}} (2005) 085005, [hep-th/0502143].
%\bibitem{ref3}B.~Basu, S.~Ghosh, Phys.Lett.{\bf{A346}}(2005) 133, [cond-mat/0503266]
\bibitem{bert}O.~Bertolami, J.~G.~Rosa, C.~M.~L.~de~Aragao, P.~Castorina, D.~Zappala, Phys. Rev. {\bf{D 72}} (2005) 025010, [hep-th/0505064].
\bibitem{rb}R.~Banerjee, B.~Dutta~Roy, S.~Samanta Phys. Rev. {\bf{D 74}} 045015 (2006) [hep-th/0605277].
\bibitem{ani}A.~Saha, [hep-th/0609195].
\end{thebibliography}
\end{document}